\documentclass[acmjacm]{acmtrans2m}
\usepackage{graphicx}

\newtheorem{theorem}{Theorem}[section]

\newtheorem{corollary}[theorem]{Corollary}

\newdef{definition}[theorem]{Definition}
\newdef{remark}[theorem]{Remark}

\def\RR{\hbox{\sf I\kern-.14em\hbox{R}}}
\def\NN{\hbox{\sf I\kern-.13em\hbox{N}}}

\newcommand{\outdeg}{\mbox{\rm outdeg}}
\newcommand{\indeg}{\mbox{\rm indeg}}
\newcommand{\core}{\mbox{\rm core}}

\newcommand{\inputpic}[3]{
\begin{figure}[!ht]
  \centering
  \includegraphics[width=#2]{#1.eps}
  \caption{\emph{#3}}
  \label{#1}
\end{figure}
}

\title{Generalized Cores}

\author{
   VLADIMIR BATAGELJ and MATJA\v{Z} ZAVER\v{S}NIK \\
   University of Ljubljana, FMF, Department of Mathematics,\\
   and IMFM Ljubljana, Department of TCS, \\
   Jadranska 19, 1000 Ljubljana, Slovenia
}

\begin{abstract}
Cores are, besides connectivity components, one among few concepts that
provides us with efficient decompositions of large graphs and networks.

In the paper a generalization of the notion of core of a graph
based on vertex property function is presented. It is shown
that for the local monotone vertex property functions the corresponding
cores can be determined in $O(m \max (\Delta, \log n))$ time.
\end{abstract}




\begin{document}

\begin{bottomstuff}
This work was supported by the Ministry of Education, Science and Sport of
Slovenia, Project J1-8532. It is a detailed version of the part of the talk
presented at \emph{Recent Trends in Graph Theory, Algebraic Combinatorics,
and Graph Algorithms}, September 2001, Bled, Slovenia.
\end{bottomstuff}

\maketitle

% ==========================================================================

\section{Cores}

The notion of core was introduced by Seidman in 1983 \cite{SEIDMAN,WaF94}.

Let $\mathbf{G}=(V,L)$ be a simple graph. $V$ is the set of \textit{vertices}
and $L$ is the set of \textit{lines} (\textit{edges} or \textit{arcs}). We
will denote $n = |V|$ and $m = |L|$. A subgraph $\mathbf{H}=(C,L|C)$ induced
by the set $C \subseteq V$ is a \textbf{$k$-\emph{core}} or a \textbf{\emph
{core of order}} $k$ iff $\forall v\in C: \deg_H(v)\geq k$ and $\mathbf{H}$
is a maximum subgraph with this property. The core of maximum order is also
called the \textbf{\emph{main} core}. The \textbf{\emph{core number}} of
vertex $v$ is the highest order of a core that contains this vertex. Since
the set $C$ determines the corresponding core $H$ we also often call it a
core.

\inputpic{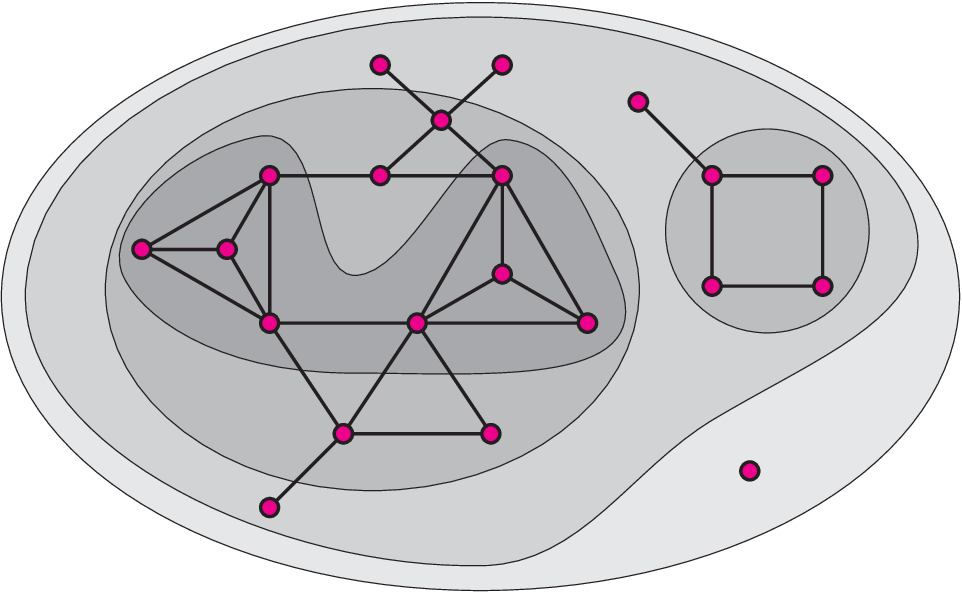}{70mm}{0, 1, 2 and 3 core}

The degree $\deg(v)$ can be the number of neighbors in an undirected graph
or in-degree, out-degree, in-degree $+$ out-degree, \ldots~ determining
different types of cores.

The cores have the following important properties:

\begin{itemize}
   \item The cores are nested: \ $i<j \ \ \Longrightarrow \ \ \mathbf{H}_j\subseteq \mathbf{H}_i$
   \item Cores are not necessarily connected subgraphs.
\end{itemize}

In this paper we present a generalization of the notion of core from degrees
to other properties of vertices.

% ==========================================================================

\section{$p$-cores}

Let $\mathbf{N} = (V, L, w )$ be a \textit{\textbf{network}}, where
$\mathbf{G} = (V, L )$ is a graph and $w : L \to \RR$ is a function assigning
values to lines. A \textit{\textbf{vertex property function}} on $\mathbf{N}$,
or a $p$ \textbf{\textit{function}} for short, is a function $p(v, U)$,
$v \in V$, $U \subseteq V$  with real values.

\bigskip\noindent
\textbf{Examples of vertex property functions:} Let $N(v)$ denotes the set of
neighbors of vertex $v$ in graph $G$, and $N(v,U) = N(v) \cap U$, $U \subseteq V$.

\begin{enumerate}
   \item $p_1(v,U) = \deg_U(v)$
   \item $p_2(v,U) = \indeg_U(v)$
   \item $p_3(v,U) = \outdeg_U(v)$
   \item $p_4(v,U) = \indeg_U(v) + \outdeg_U(v)$
   \item $p_5(v,U) = \sum_{u \in N(v,U)} w(v,u)$, where $w : L \to \RR^+_0$
   \item $p_6(v,U) = \max_{u \in N(v,U)} w(v,u)$, where $w : L \to \RR$
   \item $p_7(v,U) = $ number of cycles of length $k$ through vertex $v$
\end{enumerate}

The subgraph $\mathbf{H}=(C,L|C)$ induced by the set $C \subseteq V$ is a
$p$-\textbf{\emph{core at level}} $t \in \RR$ iff
\begin{itemize}
   \item $\forall v \in C: t \leq p(v,C)$
   \item $C$ is maximal such set.
\end{itemize}

The function $p$ is \emph{monotone} iff it has the property
$$
   C_1 \subset C_2  \Rightarrow \forall v \in V :
   ( p(v,C_1) \leq p(v,C_2 ))
$$
All among functions $p_1$ -- $p_7$ are monotone.

For monotone function the $p$-core at level $t$ can be determined by
successively deleting vertices with value of $p$ lower than $t$.

\begin{quote}
$C := V$;\\
\textbf{while} $\exists v \in C: p(v,C) < t$ \textbf{do} $C := C \setminus \{v\}$;
\end{quote}

\begin{theorem}
For monotone function $p$ the above procedure determines the $p$-core
at level $t$.
\end{theorem}
\begin{proof}
The set $C$ returned by the procedure evidently has the first property
from the $p$-core definition.

Let us also show that for monotone $p$ the result of the procedure is
independent of the order of deletions.

Suppose the contrary -- there are two different $p$-cores at level $t$,
determined by sets $C$ and $D$. The core $C$ was produced by deleting
the sequence $u_1$, $u_2$, $u_3$, \ldots, $u_p$; and $D$ by the sequence
$v_1$, $v_2$, $v_3$, \ldots, $v_q$. Assume that $D \setminus C \ne \emptyset$.
We will show that this leads to contradiction.

Take any $z \in D \setminus C$. To show that it also can be deleted we
first apply the sequence $v_1$, $v_2$, $v_3$, \ldots, $v_q$ to get $D$.
Since $z \in D \setminus C$ it appears in the sequence $u_1$, $u_2$, $u_3$,
\ldots, $u_s=z$. Let $U_0 = \emptyset$ and $U_i = U_{i-1} \cup \{ u_i \}$.
Then, since $\forall i \in 1..p : p(u_i,V \setminus U_{i-1}) < t$, we have,
by monotonicity of $p$, also $\forall i \in 1..p : p(u_i, (V \setminus D)
\setminus U_{i-1}) < t$. Therefore also all $u_i \in D \setminus C$ are
deleted -- $D \setminus C = \emptyset$ -- a contradiction.

Since the result of the procedure is uniquely defined and vertices outside
$C$ have $p$ value lower than $t$, the final set $C$ satisfies also the
second condition from the definition of $p$-core -- it is the $p$-core at
level $t$.
\end{proof}

\begin{corollary}
For monotone function $p$ the cores are nested
$$ t_1 < t_2  \Rightarrow  \mathbf{H}_{t_2} \subseteq \mathbf{H}_{t_1} $$
\end{corollary}
\begin{proof}
Follows directly from the theorem~1. Since the result is independent of
the order of deletions we first determine the $\mathbf{H}_{t_1}$. In the
following we eventually delete some additional vertices thus producing
$\mathbf{H}_{t_2}$. Therefore $\mathbf{H}_{t_2} \subseteq \mathbf{H}_{t_1}$.
\end{proof}

\bigskip\noindent
\textbf{Example of nonmonotone $p$ function:}
Consider the following $p$ function

$$
   p(v,U) = \left\{ \begin{array}{ll}
       0 & N(v,U) = \emptyset \medskip\\
       {\displaystyle \frac{1}{|N(v,U)|}\sum_{u \in N(v,U)} w(v,u) } \qquad & {\rm otherwise}
   \end{array} \right.
$$

where $w : L \to \RR^+_0$ on the network $\mathbf{N} = (V,L,w)$,
$V = \{ a, b, c, d, e, f \}$,
$$
   \begin{array}{c|ccccc}
      L & (a:b) & (b:c) & (c:d) & (b:e) & (e:f) \\ \hline
      w &   4   &   1   &   3   &   1   &   3
   \end{array}
$$
We get different results depending on whether we first delete the vertex
$b$ or $c$ (or $e$) -- see Figure~2.

\inputpic{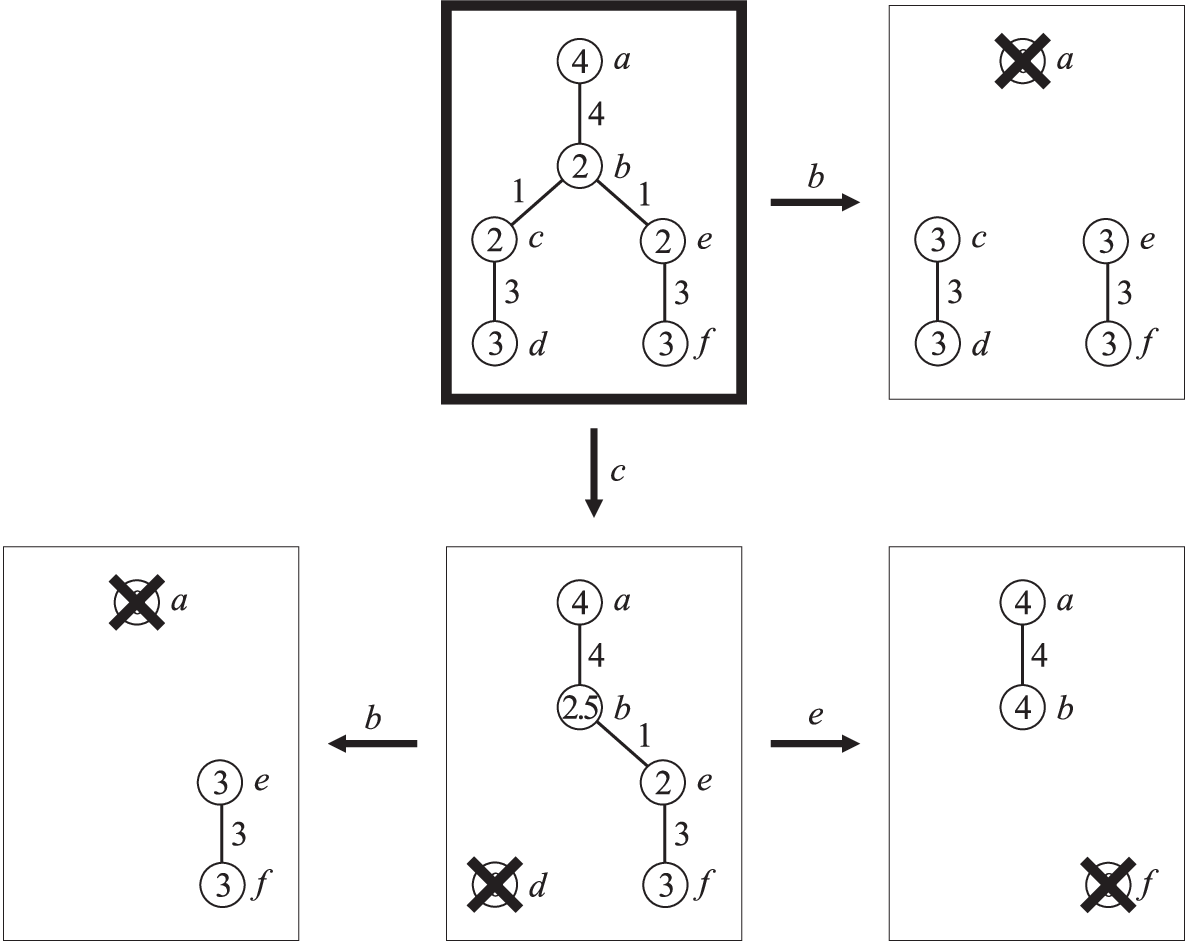}{120mm}{Nonmonotone $p$ function}

The original network is a $p$-core at level 2. Applying the algorithm to the
network we have three choices for the first vertex to be deleted: $b$, $c$
or $e$. Deleting $b$ we get, after removing the isolated vertex $a$, the
$p$-core $C_1 = \{ c, d, e, f \}$ at level 3. Note that the values of $p$
in vertices $c$ and $e$ increased from 2 to 3.

Deleting $c$ (or symmetrically $e$ -- we analyze only the first case) we get
the set $C_2 = \{ a, b, e, f \}$ at level 2 -- the value at $b$ increased to
2.5. In the next step we can delete either the vertex $b$, producing the set
$C_3 = \{ e, f \}$ at level 3, or the vertex $e$, producing the $p$-core
$C_4 = \{ a, b \}$ at level 4.

As we see, the result of the algorithm depends on the order of deletions.
The $p$-core at level 4 is not contained in the $p$-core at level 3.

% ==========================================================================

\section{Algorithms}
\subsection{Algorithm for $p$-core at level $t$}

The $p$ function is \textbf{\emph{local}} iff
$$ p(v,U) = p(v, N(v,U)) $$
The functions $p_1$ -- $p_6$ from examples are local;
$p_7$ is \textbf{not} local for $k \geq 4$.

In the following we shall assume also that for the
function $p$ there exists a constant $p_0$ such that
$$ \forall v \in V : p(v,\emptyset) = p_0 $$

For a local $p$ function an $O(m \max (\Delta, \log n))$ algorithm
for determining $p$-core at level $t$ exists (assuming that $p(v,N(v,C))$
can be computed in $O(\deg_C(v))$).

\begin{tabbing}
xxxxxxx\=xxx\=xxx\=xxx\=xxx\=xxx\=xxx\=xxx\kill
INPUT: graph $G=(V,L)$ represented by lists of neighbors and $t\in\RR$ \\
OUTPUT: $C\subseteq V$, $C$ is a $p$-core at level $t$ \\
\\
1.    \> $C := V$;\\
2.    \> \textbf{for} $v \in V$ \textbf{do} $p\lbrack v \rbrack := p(v,N(v,C))$; \\
3.    \> $build\_min\_heap(v,p)$; \\
4.    \> \textbf{while} $p\lbrack top \rbrack < t$ \textbf{do begin} \\
4.1.  \> \> $C := C \setminus \{ top \}$; \\
4.2.  \> \> \textbf{for} $v \in N(top,C)$ \textbf{do begin} \\
4.2.1.\> \> \> $p\lbrack v \rbrack := p(v,N(v,C))$; \\
4.2.2.\> \> \> $update\_heap(v,p)$; \\
\> \> \textbf{end}; \\
\> \textbf{end};
\end{tabbing}
The step 4.2.1. can often be speeded up by updating the $p\lbrack v \rbrack$.

This algorithm is straightforwardly extended to produce the hierarchy of
$p$-cores. The hierarchy is determined by the core-number assigned to each
vertex -- the highest level value of $p$-cores that contain the vertex.

\subsection{Determining the hierarchy of $p$-cores}

\begin{tabbing}
xxxxxxx\=xxx\=xxx\=xxx\=xxx\=xxx\=xxx\=xxx\kill
INPUT: graph $G=(V,L)$ represented by lists of neighbors \\
OUTPUT: table $core$ with core number for each vertex \\
\\
1.    \> $C := V$;\\
2.    \> \textbf{for} $v \in V$ \textbf{do} $p\lbrack v \rbrack := p(v,N(v,C))$; \\
3.    \> $build\_min\_heap(v,p)$; \\
4.    \> \textbf{while} $sizeof(heap)>0$ \textbf{do begin} \\
4.1.  \> \> $C := C \setminus \{ top \}$; \\
4.2.   \> \> $core\lbrack top\rbrack := p\lbrack top\rbrack$; \\
4.3.  \> \> \textbf{for} $v \in N(top,C)$ \textbf{do begin} \\
4.3.1.\> \> \> $p\lbrack v \rbrack := \max{\{p[top],p(v,N(v,C))\}}$; \\
4.3.2.\> \> \> $update\_heap(v,p)$; \\
\> \> \textbf{end}; \\
\> \textbf{end};
\end{tabbing}

Let us assume that $P$ is a maximum time needed for computing the value of
$p(v,U)$, $v \in V$, $U \subseteq V$. Then the complexity of statements 1. -- 3.
is $T_{1-3} = O(n) + O(P n) + O(n \log n ) = O(n \cdot \max (P, \log n ))$.
Let us now look at the body of the \textbf{while} loop. Since at each repetition
of the body the size of the set $C$ is decreased by 1 there are at most $n$
repetitions. Statements  4.1 and 4.2 can be implemented to run in constant
time, thus contributing $T_{4.1,4.2} = O(n)$ to the loop. In all combined repetitions
of the \textbf{while} and \textbf{for} loops each line is considered at most once.
Therefore the body of the \textbf{for} loop (statements 4.3,1 and 4.3.2)
is executed at most $m$ times -- contributing at most
$T_{4.3} = m \cdot ( P + O(\log n))$ to the \textbf{while} loop.
Often the value $p(v,N(v,C))$ can be updated in constant time -- $P = O(1)$.
Summing up all the contributions we get the total time complexity of the
algorithm $T = T_{1-3} + T_{4.1,4.2} + T_{4.3} = O(m \cdot \max (P, \log n ))$.

For a local $p$ function, for which the value of $p(v,N(v,C))$ can be computed
in\break $O(\deg_C(v))$), we have $P = O(\Delta)$.

The described algorithm is partially implemented in program for large networks analysis
\texttt{Pajek} (Slovene word for Spider) for Windows (32 bit) \cite{BMCON}.
It is freely available, for noncommercial use, at its homepage:

\texttt{http://vlado.fmf.uni-lj.si/pub/networks/pajek/}\\
A standalone implementation of the algorithm in C is available at

\texttt{http://www.educa.fmf.uni-lj.si/datana/pub/networks/cores/}\\
For the property functions $p_1$ -- $p_4$ a quicker $O(m)$ core determining
algorithm can be developed \cite{BZlin}.

% ==========================================================================

\section{Example -- Internet Connections}

\begin{table}[!tbp]
\caption{$p_5$-cores of the Routing Data Network at Different Levels.\label{levcor}}

\begin{center}
\begin{tabular}{|r|rr||r|rr|}
\hline
     $k$ &       $t$ &      $n$  &      $k$ &       $t$ &      $n$  \\
\hline
       1 &  1        &     7582  &       12 &  2048     &      490  \\
       2 &  2        &     9288  &       13 &  4096     &      314  \\
       3 &  4        &    12519  &       14 &  8192     &      153  \\
       4 &  8        &    33866  &       15 &  16384    &       48  \\
       5 &  16       &    33757  &       16 &  32768    &       44  \\
       6 &  32       &    11433  &       17 &  65536    &       11  \\
       7 &  64       &     6518  &       18 &  131072   &        9  \\
       8 &  128      &     3812  &       19 &  262144   &        0  \\
       9 &  256      &     2356  &       20 &  524288   &        2  \\
      10 &  512      &     1528  &       21 &  1048576  &        3  \\
      11 &  1024     &      918  &      &    &      \\
\hline
\end{tabular}
\end{center}
\end{table}

As an example of application of the proposed algorithm we
applied it to the routing data on the Internet network.
This network was produced
from web scanning data (May 1999) available from

\texttt{http://www.cs.bell-labs.com/who/ches/map/index.html}\\
It can be obtained also as a \texttt{\textbf{Pajek}}'s NET file from

\texttt{http://vlado.fmf.uni-lj.si/pub/networks/data/web/web.zip}\\
It has
$124\ 651$ vertices, $195\ 029$ arcs (loops were removed),
$\Delta = 151$, and average degree is $3.13$.
The arcs have as values the number of \emph{traceroute paths}
which contain the arc.

\inputpic{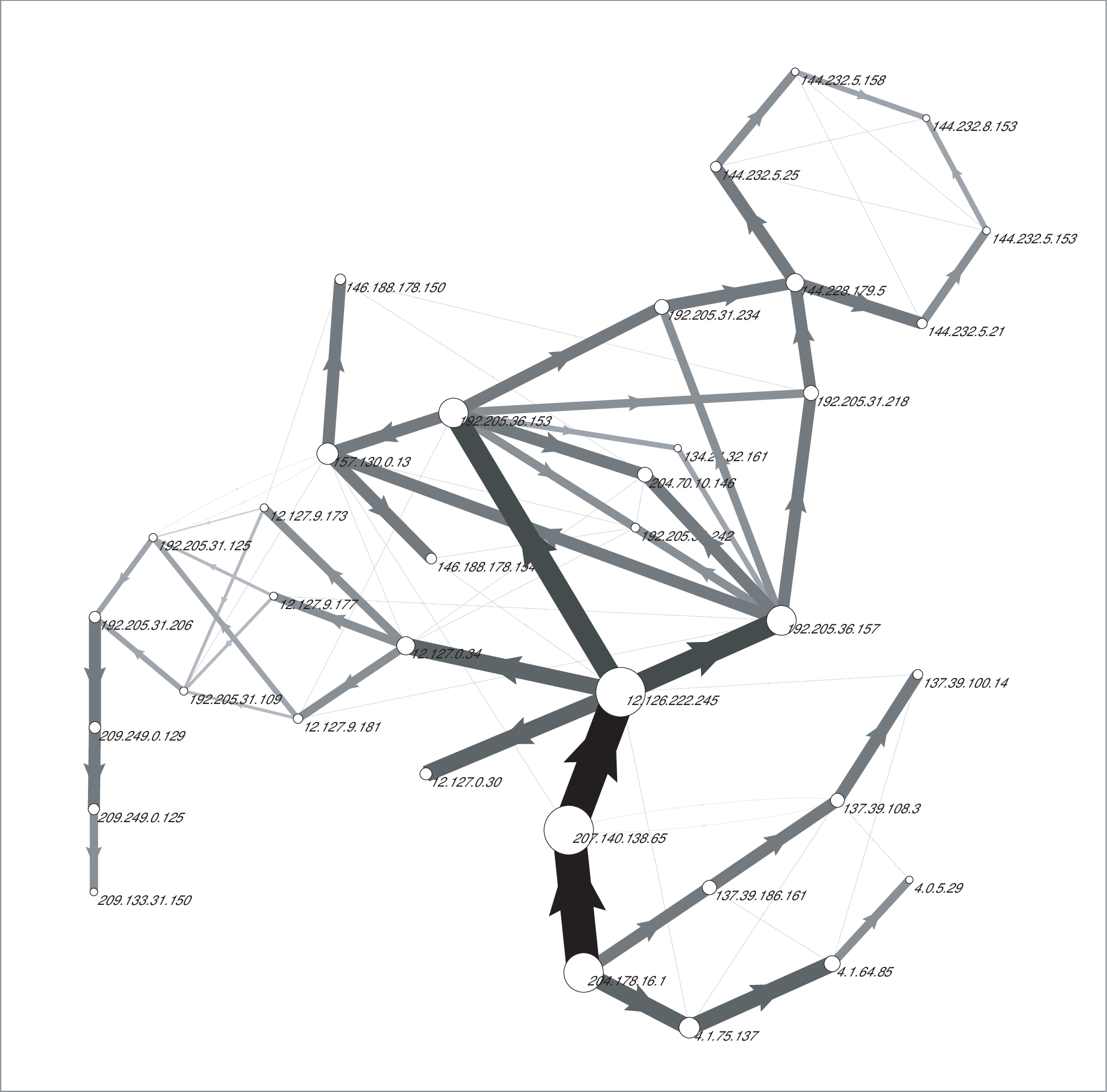}{\textwidth}{$p_5$-core of the Routing Data Network. \label{www}}

Using \texttt{\textbf{Pajek}}  implementation of the proposed algorithm on
300 MHz PC we obtained in 3 seconds the $p_5$-cores segmentation
presented in Table~\ref{levcor} -- there are $n_k$ vertices with
$p_5$-core number in the interval $(t_{k-1},t_k \rbrack$.

The program also determined the $p_5$-core number for every vertex.
Figure~\ref{www} shows a $p_5$-core at level 25000 of the Internet network --
every vertex inside this core is visited by at least 25000 traceroute paths.
In the figure the sizes of
circles representing vertices are proportional to (the square roots of)
their $p_5$-core numbers. Since the arcs values span from 1 to 626826
they can not be displayed directly. We recoded them according to the
thresholds $1000 \cdot 2^{k-1}$, $k=1, 2, 3 \ldots$. These class numbers
are represented by the thickness of the arcs.

% ==========================================================================

\section{Conclusions}

The cores, because they can be efficiently determined, are one among few concepts
that provide us with meaningful decompositions of large networks~\cite{GJ}. We expect that
different approaches to the analysis of large networks can be built on this basis.
For example, the sequence of vertices in sequential coloring can be determined
by descending order of their core numbers (combined with their degrees). We obtain
in this basis the following bound on the chromatic number of a given graph $\mathbf{G}$
$$ \chi(\mathbf{G}) \leq 1 + \core(\mathbf{G}) $$
Cores can also be used to localize the search for interesting subnetworks
in large networks \cite{BMZ,ERDOS}:

% ==========================================================================

\begin{itemize}
\item If it exists, a $k$-component is contained in a $k$-core.
\item If it exists, a $k$-clique is contained in a $k$-core.
  $\omega(\mathbf{G}) \leq \core(\mathbf{G})$.
\end{itemize}

% ==========================================================================

\bibliographystyle{acmtrans}
\bibliography{paper}

% ==========================================================================

\begin{received}
Received Month Year; revised Month Year; accepted Month Year
\end{received}
\end{document}